\documentstyle[aps,prl,preprint]{revtex}

\begin{document}
\title{Plastic vortex-creep in YBa$_2$Cu$_3$O$_{7-x}$ crystals}
\author{Y. Abulafia, A. Shaulov, Y. Wolfus, R. Prozorov, L. Burlachkov, and Y.
Yeshurun}
\address{Institute of Superconductivity, Department of Physics, \\
Bar-Ilan University, Ramat-Gan 52900, Israel}
\author{D. Majer and E. Zeldov}
\address{Department of Condensed Matter Physics, The Weizmann Institute of Science,\\
Rehovot 76100, Israel}
\author{H. W\"uhl}
\address{Forschungszentrum Karlsruhe, Institut f\"ur Technische Physik and\\
Universit\"at Karlsruhe, D-76021 Karlsruhe, Germany}
\author{V.B. Geshkenbein}
\address{Theoretische Physik, ETH Z\"urich-H\"onggerberg, CH-8093 Z\"urich,\\
Switzerland and L.D. Landau Institute for Theoretical Physics, 117940\\
Moscow, Russia}
\author{V.M. Vinokur}
\address{Argonne National Laboratory, Argonne, IL 60439, USA}
\maketitle

\begin{abstract}
Local magnetic relaxation measurements in YBa$_2$Cu$_3$O$_{7-x}$ crystals
show evidence for plastic vortex-creep associated with the motion of
dislocations in the vortex lattice. This creep mechanism governs the vortex
dynamics in a wide range of temperatures and fields below the melting line
and above the field corresponding to the peak in the ''fishtail''
magnetization. In this range the activation energy $U_{pl}$, which decreases
with field, drops below the elastic (collective) creep activation energy, $%
U_{el}$, which increases with field. A crossover in flux dynamics from
elastic to plastic creep is shown to be the origin of the fishtail in YBa$_2$%
Cu$_3$O$_{7-x}$.

PACS numbers: 74.60.Ge, 74.72.Bk
\end{abstract}

Magnetic relaxation in high-temperature superconductors has been
traditionally described in terms of collective vortex-creep based on the
concept of elastic motion of the vortex lattice \cite{Blatter,Brandt-rev}.
This approach successfully explained wide range of vortex dynamics
phenomena. On the other hand, under certain conditions effects of vortex
lattice {\em plasticity} may become the dominant factor that determines the
vortex dynamics \cite{two-D}. For example, numerical simulations of strongly
pinned vortex system reveal vortex motion dominated by plastic deformations 
\cite{numerics1}. Experimentally, effects of plastic vortex behavior were
observed in transport measurements in close vicinity of the melting
transition where the pre-melting softening of the lattice enhances the role
of spatial inhomogeneities, resulting in the tearing of the vortex lattice
under applied currents \cite{transport,Kwok}. In this paper we demonstrate
that plastic deformations dominate vortex-lattice motion in the creep regime
in YBa$_2$Cu$_3$O$_{7-x}$ (YBCO) crystals far below the melting transition,
in the region that was believed to be governed by elastic motion. We find
that this plastic motion governs the flux creep in YBCO crystals at elevated
temperatures at fields above the characteristic field $B_p$ corresponding to
the peak in the 'fishtail' magnetization \cite
{Daeumling,Krusin,Gordeev,Yeshurun,Klein,Kupfer,Zhukov,Wen}, and it affects
the corresponding shape of the magnetization curves.

The plastic vortex motion can be classified into three main categories. i)
Vortex channeling along easy paths in the pinning relief in between rather
stationary vortex-lattice islands. Such behavior was observed in numerical
simulations in presence of very strong pinning \cite{numerics1}. ii) Vortex
motion that resembles ice floe in which large pieces of vortex-lattice slide
with respect to each other. As was simulated numerically \cite{numerics2},
these two modes of dynamics are believed to be the cause of the highly
unstable character of resistivity in the close vicinity of the melting
transition \cite{transport,Kwok}. iii) Dislocation mediated plastic creep of
vortex lattice similar to diffusion of dislocations in atomic solids \cite
{HL}. This type of vortex behavior was considered previously \cite{FGLV,GLFV}
without an experimental evidence. The observations described below strongly
suggest that this mode of plastic vortex creep governs magnetic relaxation
over a substantial part of the YBCO magnetic phase diagram.

Local magnetic relaxation measurements were performed on a $1.2\times
0.5\times 0.3~mm^3$ single crystal of YBCO ($T_c\simeq 91\,K$) using an
array of microscopic GaAs/AlGaAs Hall sensors with $30\times 30\ \mu m^2$
active area and sensitivity better than $0.1\,G$. The probes detect the
component $B_z$ of the field normal to the surface of the crystal.
Temperature stability and resolution was better then $0.01~K$.

After zero-field-cooling ({\it zfc}) the sample from above $T_c$ to the
measurement temperature $T$ we measured the full hysteresis loops for all
the probes. The first field for full penetration $H^{*}$ was measured
directly by the probe at the center of the sample. After repeating the {\it %
zfc} process, a {\it dc} field $H$ was applied parallel to the $c$-axis and
the local induction $B_z$ was measured at different locations as a function
of the time $t$ for an hour. These relaxation measurements were repeated
after the field was increased by a step $\Delta H>2H^{*}$ up to the
irreversibility field $H_{irr}$ or the maximum field of the experiment ($%
1.6~T$).

The inset to Figure 1 shows typical hysteresis loops, $B_z^{(i)}-H$ vs. $H$
at $T=85~K$ for four probes ($i=5,6,7,8$ located at $70$, $130$, $190$, and $%
250$ $\mu m$ from the edge towards the center). Each probe exhibits a clear
fishtail behavior with a maximum in local magnetization at field $B_p\simeq
0.4~T$. The width of the loop is largest in the center of the crystal and
decreases towards the edges, as expected from basic considerations based on
a modified Bean model \cite{Prozorov}.

In Figure 1 we show the time evolution of the gradient $\left(
B_z^{(6)}-B_z^{(7)}\right) /\Delta x$ between $t_1=8$ sec and $t_2=3600~\sec 
$, as a function of the applied field. In our geometry (aspect ratio $=3/5$)
this gradient is proportional to the persistent current density $J${\bf \ }%
that can be readily evaluated using sensors $6$ and $7$ which are located
not too close to the center or the edge of the crystal \cite
{Abulafia2,strip,Abulafia1}. Note that the position of the fishtail peak $%
B_p $ shifts from $0.42~T$ to $0.34~T$ during the relaxation. The total
relaxation of the current, $\Delta J$, during the time window of the
measurement is large{\sf \ }for $B<B_p$ and it increases to even larger
values above $B_p$. Furthermore, the relative change of the persistent
current, $\Delta J/J$, is also large (e.g. $\Delta J/J\approx 0.8$ at $H=0.9$
$T$), implying that $J\ll J_c$ on both sides of the fishtail peak. These
strong relaxations imply that dynamic effects determine the shape of $J(B)$
and rule out the possibility \cite{Daeumling,Krusin,Kupfer} that either
branch of the fishtail is determined by the critical current density $J_c(B)$%
.

Knowledge of the time and spatial field-derivatives, $\partial B_z/\partial
t $ and $\partial B_z/\partial x$, enable direct determination \cite
{Abulafia1} of the activation energy $U(J,B)$ associated with the flux creep
by using the diffusion equation: $\partial B_z/\partial t=-\partial
/\partial x(B_zv)$, where the effective vortex velocity $v$ is proportional
to $\exp (-U/kT)$. Typical $U$ vs. $J$ data, at $85~K$, are shown in Fig.~2
for different fields, in the range of $0.05$ $T$ to $0.8$ $T$. This figure
exhibits a dramatic crossover in the slope $\left| dU/dJ\right| $ around $%
B_p=0.4$ $T$. In order to quantify this crossover, we start by using the
prediction of the collective creep theory for $J\ll J_c$ \cite{Blatter,FGLV}:

\begin{equation}
U(B,J)=U_0(B)(J_c/J)^\mu \propto B^\nu J^{-\mu },  \label{eq-U(B,j)}
\end{equation}
where the positive critical exponents $\nu $ and $\mu $ depend on the
specific pinning regime. We note that the range of the experimentally
accessible $U$-values in Fig.~2 is almost independent of the field \cite
{U(j)}. This implies $J\propto B^{\nu /\mu }$, i.e. $J$ {\em grows with field%
} for $J\ll J_c$. Obviously, the collective creep dynamics cannot explain
the decrease of $J$ with $B$ observed above $B_p$.

The inset to Fig.~2 shows the $\mu $-values obtained by fitting the $U(J)$
data, using Eq.~\ref{eq-U(B,j)}. At low fields $\mu \simeq 1$ and it then
increases to $\mu \simeq 2$ just below $B_p$. This confirms \cite{Krusin}
that below the peak the relaxation is well described by the collective creep
theory. The latter predicts $\mu =1$ in the intermediate bundle regime and $%
\mu =5/2$ in the small bundle regime \cite{bundle}. However, above the peak, 
$\mu $ drops sharply to values below $0.2$. Within the collective creep
theory this would imply an inconceivable crossover to a single vortex regime
($\mu =1/7$) which is expected only for low fields and high values of $J$ 
\cite{Blatter}. Thus the $\mu $-values above $B_p$ are inconsistent with the
collective creep theory.

The failure of the collective creep theory in explaining the data above $B_p$
can be further demonstrated by analyzing the exponent $\nu $ in Eq.~(\ref
{eq-U(B,j)}). Separation of variables in this equation implies that a smooth 
$U(J)$ function can be obtained by proper field-scaling of the data sets at
various fields. However, as demonstrated in Fig. 3 we need {\em two}
exponents of{\em \ opposite signs}, $\nu \cong 0.7$ and $\nu \cong -1.2$, to
scale the data of Fig.~2 for fields below and above $B_p$, respectively. In
the collective creep theory $U$ increases with $B$, thus the{\em \ negative} 
$\nu $-value is inconsistent with this theory.

We now show that the experimental data above $B_p$ can be well explained by
a plastic creep model based on dislocation mediated motion of vortices
similar to diffusion of dislocations in atomic solids \cite{HL}. The
activation energy $U_{pl}^0$ at $J=0$ for the motion of a dislocation in the
vortex lattice can be estimated as \cite{GLFV}:

\begin{equation}
U_{pl}^0(B)\simeq \varepsilon \varepsilon _0a\propto 1/\sqrt{B}
\label{eq-Upl}
\end{equation}
where $\varepsilon _0$ is the vortex line tension, $\varepsilon =\sqrt{%
m_{ab}/m_c}$ is the anisotropy parameter, and $a\simeq \sqrt{\phi _0/B}$ is
the mean intervortex distance. This estimation assumes a formation of a
dislocation semi-loop between two valleys separated by a distance $a$ \cite
{HL}. One notices that $U_{pl}^0$ {\em decreases} with field in contrast to
the collective creep activation energies $U_{el}$, which {\em increases}
with field, see Eq.~(\ref{eq-U(B,j)}). Of course, the creep process is
governed by the smaller between $U_{el}$ and $U_{pl}$. Thus, at low fields
where $U_{pl}>U_{el}$, the latter controls the flux dynamics. But, as $B$
increases and $U_{pl}$ becomes less than $U_{el}$, a crossover to the
plastic creep regime is expected.

The values of $U_{pl}^0(B)$ can be extracted from the measured $U(J)$ curves
of Fig.~2 assuming an expression for $U_{pl}(J)$ taken from the dislocation
theory \cite{HL}, substituting the current density for the strain: 
\begin{equation}
U_{pl}(J)=U_{pl}^0\left( 1-\sqrt{J/J_c^{pl}}\right) ,  \label{eq-Upl(j)}
\end{equation}
where $J_c^{pl}$ is the critical current which corresponds to the plastic
motion. The derived $U_{pl}^0$ values are shown in the inset of Fig.~3 as a
function of $B$. The solid line in the inset is a power-law fit to the
experimental data, $U_{pl}^0\propto B^{-0.7}$. Clearly, $U_{pl}^0$ decreases
with the field, although with an exponent $-0.7$ rather than the expected $%
-0.5$. As we show below, the same exponent ($-0.7$) is also involved in
determining the temperature dependence of $B_p$.

The fishtail peak location $B_p$ can be determined from the condition $%
U_{el}=U_{pl}$ for the same $J$. Note that $U_{pl}\cong U_{pl}^0$ since $%
J\ll J_c^{pl}$. Using the logarithmic solution of the flux diffusion
equation \cite{Blatter} $U_{el}=kT\ln (t/t_0)$ together with Eq.~(\ref
{eq-Upl}), we get $B_p\propto 1/\ln ^2(t/t_0)$, i.e. the peak position
should shift with time towards low fields, as observed in Fig.~1. Similarly,
for the temperature dependence of $B_p$ one obtains: 
\begin{equation}
B_p\propto \varepsilon _0^2\propto 1/\lambda ^4\propto \left(
1-(T/T_c)^4\right) ^2.  \label{eq-HpeakT}
\end{equation}
A fit of this expression to the experimental data yields a modest agreement.
However, as noted above, in fact $U_{pl}^0\propto B^{-0.7}$, thus $%
B_p(T)\propto \varepsilon _0^{1/0.7}$, i.e. $B_p\propto \left(
1-(T/T_c)^4\right) ^{1.4}$. Indeed, a perfect fit ($B_p$, solid line in Fig.
4) is obtained with this expression. Note that a similar variation of the
exponent was found in some experiments \cite{melting} for the melting line $%
B_m(T)\propto \left( 1-(T/T_c)\right) ^{1.4}$, while the theory \cite
{Blatter} predicted a critical exponent of $2$. This interesting similarity
may support previous claims \cite{Kwok} that the plastic motion of defects
in the vortex lattice is a precursor to the melting transition.

In Figure 4 we show that the plastic creep regime in the $B-T$ phase diagram
(shaded area) covers a substantial area between the $B_p(T)$ line found in
this experiment and the melting line $B_m(T)$ described in Ref. \cite
{melting}.

In conclusion, our data clearly indicate two different flux-creep mechanisms
above and below the peak in the magnetization curves. In particular, flux
creep with activation energy {\em decreasing} with field plays an important
role above the peak. The data in this regime cannot be explained in terms of
the traditional collective creep theory based on the concept of elastic
motion of the vortex lattice. On the other hand, the data show good
agreement with the dislocation mediated mechanism of plastic creep analogous
to plasticity in atomic solids. This observation leads also to the
conclusion that the origin of the fishtail in YBCO crystals is a crossover
from elastic to plastic creep. The predictions of this model for the time
and temperature dependence of the location of the peak are well confirmed in
the experiments.

We thank H. Shtrikman for growing the GaAs heterostructures, and acknowledge
useful discussions with M. Konczykowski and P. Kes. This work was supported
in part by the Israel Academy of Science and Humanities, the Heinrich Hertz
Minerva Center for High Temperature Superconductivity, and the DG XII,
Commission of the European Communities, administered by the Israeli Ministry
of Science and the Arts. Y.Y. and E.Z. acknowledge support of the
U.S.A.-Israel Binational Science Foundation. A.S. and E. Z. acknowledge
support from the France-Israel cooperation program AFIRST. V.B.G. is
grateful to ITP at UCSB for the support via grant PHY94-07194. V.M.V.
acknowledges support from the US Department of Energy, BES - Material
Sciences, under contract no. W-31-109-ENG-38.\ The visits of V.B.G. and
V.M.V. to Bar-Ilan were supported by the Rich Foundation.

\begin{center}
{\sc FIGURE CAPTIONS}
\end{center}

\begin{itemize}
\item[Fig. 1]  Relaxation of persistent current $J$ calculated from the
field gradients for different applied fields at $T=85~K$. Note the shift in
the peak position $B_p$ during the experimental time window $8-3600$ sec as
noted by the arrows. Also note the increase in the relaxation rate above $%
B_p $. Inset: Local hysteresis loops for four Hall probes vs. applied field $%
H$ at $T=85\,K$. The width of the loops increases from the edge of the
sample towards the center.

\item[Fig. 2]  $U$ vs. $J$ for the indicated fields ($0.05$ $T-0.8$ $T$) at $%
T=85\,K$. The lines are guide to the eye. Note the change in the slope $%
\partial U/\partial j$ above and below the peak field $B_p\simeq 0.4~T$.
Inset: The critical exponent $\mu $, see Eq.~(\ref{eq-U(B,j)}), as a
function of field, at $T=85\,K$.

\item[Fig. 3]  Scaling of $U(J,B)$-curves below and above $B_p$ at $T=85\,K$%
. Above the peak $U\propto B^{0.7}$ and below the peak $U\propto B^{-1.2}$.
The inset shows the derived $U_{pl}^0$ vs. $B$. The solid line is a fit to $%
U_{pl}\propto B^{-0.7}$.

\item[Fig. 4]  Vortex-creep phase diagram for YBCO. The plastic creep regime
is limited between $B_p(T)$ (solid line) and the melting line $B_m(T)$
(dashed line, taken from \cite{melting}). Data points in the $B_p(T)$ curve
were determined in this experiment at $t_1=8~\sec $. The solid line is a fit
to $B_p\propto \left( 1-(T/T_c)^4\right) ^{1.4}$.
\end{itemize}


\begin{references}
\bibitem{U(j)}  This is consistent with the logarithmic solution of the flux
diffusion equation \cite{Blatter} $U(t)=kT\ln (t/t_0)$ for the fixed
experimental time window $t_1<t<t_2$, since the only field dependent
parameter in $U(t)$ is the time scale $t_0$ under the logarithm.
\bibitem{bundle}  The intermediate bundle regime ($\mu \approx 1$)
corresponds to lower fields where the characteristic energy $U_0(B)$ is
small, and thus for a constant $U$, the ratio $J_c/J$ is large, see Eq.~\ref
{eq-U(B,j)}. This implies that $J$ relaxes down deeply below $J_c$, thus
reaching this regime, see \cite{Blatter}. As the field increases $U_0(B)$
increases thus $J_c/J$ decreases and $J$ enters the small bundle regime with 
$\mu \approx 5/2$.
\end{references}
\end{document}